\documentclass[journal,twoside,web]{ieeecolor}
\usepackage{generic}
\usepackage{cite}
\usepackage{amsmath,amssymb,amsfonts}
\usepackage{graphicx}
\usepackage{textcomp}
\usepackage{epstopdf}
\usepackage{siunitx}
\usepackage{amssymb}
\usepackage{subfig}
\usepackage{color}
\usepackage{booktabs}
\usepackage {tabularx}
\def\BibTeX{{\rm B\kern-.05em{\sc i\kern-.025em b}\kern-.08em
		T\kern-.1667em\lower.7ex\hbox{E}\kern-.125emX}}
\begin{document}
	
	\title{Negative Capacitance DG Junctionless FETs: A Charge-based Modeling Investigation of Swing, Overdrive and Short Channel Effect}
	
	\author{Amin Rassekh, Jean-Michel Sallese, Farzan Jazaeri, Morteza Fathipour and  Adrian M. Ionescu
		\thanks{Amin Rassekh and Adrian M. Ionescu are with Nanoelectronic Devices Laboratory, École Polytechnique Fédérale de Lausanne (EPFL), Switzerland (e-mail: amin.rassekh@epfl.ch). Jean-Michel Sallese, and Farzan Jazaeri are with the Electron Device Modeling and Technology Laboratory (EDLAB) of the École Polytechnique Fédérale de Lausanne, Switzerland. Morteza Fathipour is with device simulation and modeling Laboratory, department of electrical and computer engineering, University of Tehran, Iran. received XXXX XX, 2020.}\vspace{-1.3cm}}
	\maketitle
	\begin{abstract}
		In this paper,  an analytical predictive model of the negative capacitance (NC) effect in symmetric long channel double-gate junctionless transistor is proposed based on a charge-based model. In particular, we have investigated the effect of the thickness of the ferroelectric on the I-V characteristics. Importantly, our model predicts that the negative capacitance minimizes short channel effects and enhances current overdrive, enabling both low power operation and more efficient transistor size scaling, while the effect on reducing subthreshold slope shows systematic improvement, with subthermionic subthreshold slope values at high current levels. Our predictive results in a long channel junctionless with NC show an improvement in ON current by a factor of 6 in comparison to junctionless FET. The set of equations can be used as a basis to explore how such a technology booster and its scaling will impact the main figures of merit of the device in terms of power performances and gives a clear understanding of the device physics. The validity of the analytical model is confirmed by extensive comparisons with numerical TCAD simulations in all regions of operation, from deep depletion to accumulation and from linear to saturation.
		
	\end{abstract}

	
	\section{Introduction}
	\IEEEPARstart{A}{dvanced} aggressive scaling of conventional metal-oxide-semiconductor field-effect transistors (MOSFET) requires the use of advanced processing with multiple additive technology boosters, such as strain, high-k dielectrics with metal gate stacks, shallow junctions and the replacement of the silicon channel with materials having higher carrier mobility \cite{collaert2018high,chau2019process}. Even more sophisticated techniques for locally controlling the strain have been proposed in various research works \cite{moselund2010high}. Significant efforts are needed for the junction and contact engineering in such advanced MOSFETs. A lot of effort has been recently dedicated to the so-called steep-slope transistors \cite{ionescu2017beyond} but their maturity is still far from being adopted by the nanoelectronics industry. In an attempt to remove all the limitations related to the junction engineering at nanoscale, the concept of junctionless field-effect transistors (JLFET) has been proposed, where the conduction in a very thin, highly doped semiconductor film is controlled by a gate field effect. Because of the absence of source and drain junctions, junctionless transistors are free from steps to create ultra-steep junctions and high thermal annealing for S/D dopant activation, which is a big advantage for scaling and cost reduction at the nanoscale \cite{lee2009junctionless}. 
	
	However, the scaling of MOS devices must cope with issues such as increased power consumption and degraded off-state current \cite{takagi2007carrier} upon reduction of the power supply to mitigate power consumption. The main parameter limiting the power supply voltage scaling in MOS devices is the intrinsic limit of \SI{60}{mV/dec} at \SI{300}{K} of the subthreshold swing (SS). To overcome this, it was proposed to add to the conventional insulator of the gate oxide a ferroelectric material of a given thickness that will create an effective negative capacitance, resulting in a reduction of the transistor body factor below unity. This would lead to SS values lower than \SI{60}{mV/dec} \cite{rusu2010metal, ionescu2017sub, saeidi2017negative, saeidi2019negative, khan2015negative, appleby2014experimental, salahuddin2008use, ionescu2010hysteretic, salvatore2008demonstration}. Having negative capacitance means that a given charge density in the channel can be achieved with a lower gate voltage.  
	
	In this work, we investigated by calibrated modeling and simulations, the effect of negative capacitance on the characteristics of junctionless transistors. The phenomena of a ferroelectric material have been modeled by Ginzburg and Devonshire which is based on the Landau theory of phase transitions \cite{landau1937theory}. Landau theory can predict the behavior of ferroelectric material. To the best of our knowledge, it is the most common approach to the study of negative capacitance effects with ferroelectrics \cite{ionescu2017sub, saeidi2017negative, saeidi2019negative, khan2015negative, appleby2014experimental, salahuddin2008use, ionescu2010hysteretic, landau1937theory, hoffmann2019unveiling, saeidi2016double}. The simulation and modeling study of negative capacitance in the MOSFETs has been investigated in the literature \cite{jiang2015carrier,mehta2018impact,choi2019device,gupta2020negative,mehta2016modeling}. Also, some studies have been done to investigate the short channel effect in inversion-mode NC FETs, a majority of them being based on detailed simulations \cite{ota2016fully,dong2017simple,seo2017analysis}. However, none of them is an analytical charge-based model and there is no evidence of modeling and investigating the short channel effect in NC JLFETs based on solving 2D Poisson-Boltzmann relationships.
			\begin{figure*}[t]
		\centering
		\includegraphics[width=2\columnwidth]{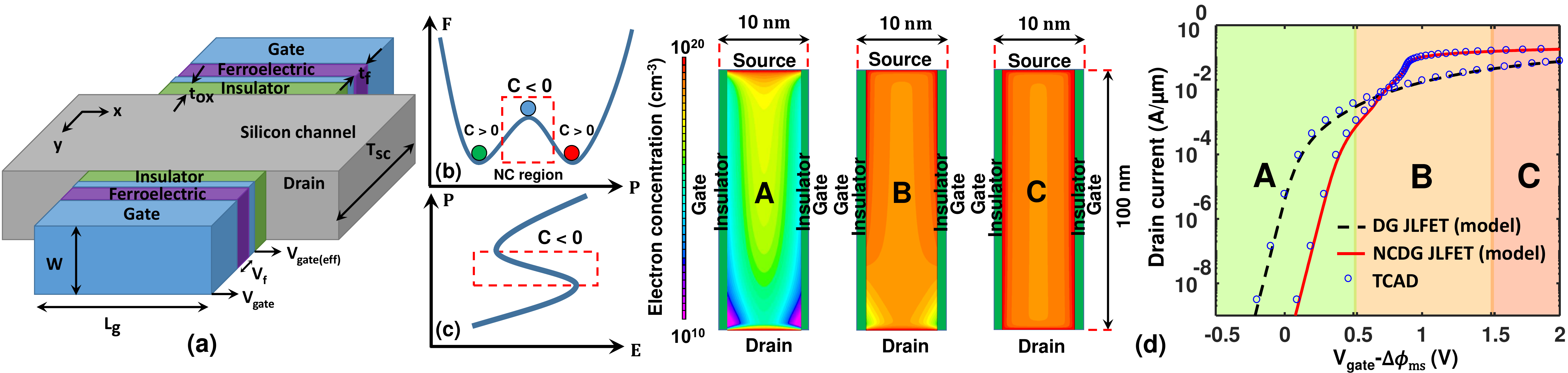}
		\caption{(a) 3-D Schematic view of a double-gate JLFET with negative capacitance. (b) The double-well free energy in ferroelectric versus the electric polarization ($P$) from the Landau theory of ferroelectrics. (c) Polarization of ferroelectric as a function of the electric field. The red rectangular in (b) and (c) denotes the region of negative capacitance. (d) Drain current versus the applied gate voltage and electron concentration corresponds to depletion, accumulation and hybrid channel mode of a junctionless transistor at $ V_{DS}$ = \SI{1}{V}.}
		\label{Fig1}\vspace{-0.5cm}
	\end{figure*}
		
	Thus, in order to take account the ferroelectric in junctionless transistors in a simple and compact model approach,  in this work, we propose analytical and explicit relationships taking into account the negative capacitance effect in Double-Gate JLFET (NCDG JLFET) and evidences an amplification of the current-voltage dependence with respect to the ferroelectric thickness. This model relies on the charge-based approach developed in \cite{sallese2011charge, jazaeri_sallese_2018, jazaeri2013analytical,rassekh2019modeling, jazaeri2014generalized, shalchian2018charge, jazaeri2014modeling}. This approach will be validated with technology computer-aided design (TCAD) simulations in all regions of operation from deep depletion to accumulation and linear to saturation. The effect of the negative capacitance on DIBL (Drain Induced Barrier Lowering) based on the 2D Poisson-Boltzmann relationship is also addressed.	
	\vspace{-0.2cm}
	\section{Modeling Approach}
	Three-dimensional schematic of the structure of the symmetric double-gate junctionless transistor with the ferroelectric layer is shown in Fig. \ref{Fig1}.a. In this structure, an intermediate metallic layer is considered between the insulator and the ferroelectric material. This inner metal film in the gate stack architecture imposes a uniform electric field inside the ferroelectric \cite{rusu2010metal}. Also, in this paper, we assume a single-domain ferroelectric film for simplicity. Hence, the single-domain Landau-Khalatnikov theory will be eligible for modeling the ferroelectric dielectric. It has been reported in \cite{hoffmann2017modeling} that in a case in which we have a multi-domain ferroelectric film, using an inner metal cannot stabilize negative capacitance. However, a lot of empirical works have been done with inner metal and they observe the NC effect \cite{rusu2010metal, ionescu2017sub, saeidi2017negative, saeidi2019negative, khan2015negative, appleby2014experimental}. The reason could be that in reality, one of the domains might be dominant and the worst case is that only the biggest domain will stabilize which is similar to a single-domain case. We call the potential of this inner metal layer $V_{gate(eff)}$ and the potential of the real gate, $V_{gate}$. Hence the potential across the ferroelectric can be expressed as	
	\begin{equation} \label {1}
	V_{f}=V_{gate}-V_{gate(eff)}.
	\end{equation}
	According to the Landau theory, the Gibbs free energy $F(J/m)$ of the ferroelectric material, respect to the polarization $P$ is
	\begin{equation} \label {2}
	F=\alpha P^2+\beta P^4+\gamma P^6-\vec{E}\vec{P},
	\end{equation}
	where $\alpha $, $\beta $, $\gamma $ are ferroelectric material constants and $\vec{E}$ is the electric field in the ferroelectric. The $F$-$P$ curve has two minima as shown in Fig. \ref{Fig1}.b. These represent the two counter stable states in the ferroelectric material ($\pm P$) which can be switched by applying an external electric field (note that since the electric field and polarization are uniform and aligned with the y-axis, we will consider them as scalars). Then we have $Q$=$P$ and $V_f$=$E$$t_f$ \cite{salahuddin2008use}, where $Q$ is the charge density of the ferroelectric (per unit area) and $t_f$ is the ferroelectric thickness. The derivation of $U_F=Ft_f$ with respect to $Q$ gives (see Fig. \ref{Fig1}.c) 
	\begin{equation} \label {3}
	\frac{\partial U_F}{\partial Q}=2\alpha t_fQ+4\beta t_fQ^3+6\gamma t_fQ^5-V_f-Q\frac{\partial V_f}{\partial Q},
	\end{equation}
	The capacitance is related to the slope of the $P$-$E$ curve as follows (see the Appendix)
	\begin{equation} \label {222}
	C=\frac{1}{t_f}\left(\epsilon_0+\frac{dP}{dE}\right),
	\end{equation}
	thus, as illustrated in Fig. \ref{Fig1}.b and c, there is a region where the capacitance is negative (red rectangular). The negative capacitance region is naturally unstable but can be stabilized when combined with an ordinary capacitor in series \cite{salahuddin2008use}. The free energy of an ordinary capacitor ($C_D$) which is in series with the ferroelectric is given by \cite{hoffmann2017modeling}
	\begin{equation} \label {223}
	U_D=\frac{Q^2}{2C_D}-QV_D,
	\end{equation}
	Where $V_D$ is the potential across the ordinary capacitor. Therefore, the minimum of the total free energy ($U=U_D+U_F$) happens when $\partial U/\partial Q=0$
	\begin{equation} \label {224}
	V_f+V_D=\frac{Q}{C_D}+2\alpha t_fQ+4\beta t_fQ^3+6\gamma t_fQ^5-Q\frac{\partial V_G}{\partial Q},
	\end{equation}
	Where $V_G=V_f+V_D$ is a constant voltage across the ferroelectric-insulator stack and we also know $V_D=Q/C_D$. Hence, NC happens for a specific $Q$-$V_f$ relationship
	\begin{equation} \label {4}
	V_f=2\alpha t_fQ+4\beta t_fQ^3+6\gamma t_fQ^5.
	\end{equation}
	
	From the charge-based model in \cite{sallese2011charge, jazaeri_sallese_2018}, what we know is the relationship between the $V_{gate(eff)}$ and the charge density in the channel. In addition, from (\ref{1}) and (\ref{4}), we also know how $V_{gate}$ and $V_{gate(eff)}$ are interrelated. These sets of relations will be now combined to simulate the device characteristics with respect to the external voltages (gate, source, and drain).
	\begin{table}
		\centering
		\caption{Device parameters used for TCAD and Model.}
		\label{tbl1}       
		\begin{tabular} {p{3cm}>{\centering}p{2.5cm}p{1.5cm}}
			\textbf{Device Parameter} & \textbf{Symbol} & \textbf{Value}  \\
			\noalign{\smallskip}
			\hline\hline
			\noalign{\smallskip}
			Gate oxide thickness & $t_{ox}$ & \SI{1}{nm} \\
			Channel length & $L_g$ & \SI{100}{nm} \\
			Channel Width & $W$ & \SI{1}{\micro m} \\
			Channel thickness & $T_{sc}$ & \SI{10}{nm} \\
			Doping concentration & $ N_D$ & \SI{e19}{cm^{-3}} \\
			Ferroelectric thickness & $t_{f}$ & \SI{0}-\SI{12}{nm} \\
			Remanent polarization & $P_r$ & \SI{17}{\micro C/cm^{2}} \\
			Coercive field & $E_c$ & \SI{1.2}{MV/cm} \\
			\noalign{\smallskip}
			\hline
		\end{tabular}\vspace{-0.5cm}
	\end{table}
\vspace{-0.2cm}
	\subsection{Recalling JLFET Core Equations} 
	We consider an n-type long-channel symmetric double-gate JLFET with ferroelectric material (as shown in Fig. \ref{Fig1}.a) with a doping density $N_D$, a channel length, thickness and width $L_g$, $T_{sc}$ and $W$  respectively.  Gate oxide and ferroelectric film thicknesses are $t_{ox} $ and $t_f$. Device parameters are listed in Table \ref{tbl1}. When the JLT is OFF, the channel becomes depleted of majority carriers and results in negligible current conduction (See A in Fig. \ref{Fig1}.d). By increasing the gate voltage carriers can pass through the channel and the electron concentration increases (See B in Fig. \ref{Fig1}.d). Finally, when the JLT is ON a high electron concentration in the channel appears, facilitating current conduction between source and drain (See C in Fig. \ref{Fig1}.d) \cite{rassekh2020single}. According to the derivation of the charge-based model for double-gate symmetric JLFETs developed in \cite{sallese2011charge, jazaeri_sallese_2018, jazaeri2013analytical, rassekh2019modeling}, we have the two following relationships which link The total charge in the semiconductor, $ Q_{sc} $ and the effective gate voltage ($V_{gate(eff)}$):
	\begin{equation} \label {7}
	\begin{split}
	\left(\frac{Q_{sc}}{2\epsilon_{si}}\right)^2&\!\!\!=\frac{2qn_{i}U_{T}}{\epsilon_{si}}\Biggl\{\!\exp\left({\frac{\psi_{0}-V_{ch}}{U_{T}}}\right)\!\!\left[\exp\left({\frac{\psi_{s}-\psi_{0}}{U_{T}}}\right)\!-1\right]\\
	&-\frac{N_D}{n_i}\left(\frac{\psi_{s}-\psi_{0}}{U_{T}}\right)\Biggr\},
	\end{split}
	\end{equation}
	\begin{equation} \label {8}
	Q_{sc}=-2C_{ox}\left(V_{gate(eff)}-\Delta\phi_{ms}-\psi_{s}\right),
	\end{equation}
	where $\psi_{s}$ and $\psi_{0}$ are the surface potential and the center potential respectively, $n_i$ is the intrinsic carrier concentration, $\epsilon_{si}$ is the permittivity of silicon, $V_{ch}$ is the quasi-Fermi potential, $U_{T}=k_BT/q$ is the thermal voltage, $C_{ox}$ is the capacitance of the insulator, $ \Delta\phi_{ms} $ denotes the difference between the work function of metal and the work function of the intrinsic semiconductor, other symbols having their usual meaning.
	\subsubsection{Depletion Mode}
	In depletion mode, the potential at the center of the semiconductor channel is higher than the surface potential, and the net charge density in the semiconductor is positive ($ Q_{sc}\geq0 $). Therefore, the exponential term in (\ref{7}) is negligible. By manipulating \cite{sallese2011charge, jazaeri_sallese_2018, jazaeri2013analytical, rassekh2019modeling}, the effective gate potential in the depletion mode with respect to the total charge density is as follows
	\begin{equation} \label {10}
	\begin{split}
	V_{gate(eff)}&=\Delta\phi_{ms}+V_{ch}-\frac{Q_{sc}}{2C_{ox}}+U_{T}\ln\left(\frac{N_D}{ni}\right)\\
	&+U_{T}\ln\left[1-\left(\frac{Q_{sc}}{Q_f}\right)^2\right]-\frac{Q_{sc}^2}{8C_{sc}Q_f},
	\end{split}
	\end{equation}
	where $Q_f=qN_DT_{sc}$ is the fixed charge in the channel and $C_{sc}=\epsilon_{si}/T_{sc}$.
	\subsubsection{Accumulation Mode}
	Under accumulation mode, the last term in (\ref{7}) is always smaller than the exponential term, which leads to a negative charge density in the semiconductor ($Q_{sc}\leq0$). In addition, in accumulation, the center potential remains close to the value it takes at the flat-band condition $\psi_{0}\approx V_{ch}+U_T\ln(N_D/n_i)$ \cite{sallese2011charge}. Therefore, the effective gate voltage in accumulation mode becomes
	\begin{equation} \label {12}
	\begin{split}
	V_{gate(eff)}&=\Delta\phi_{ms}+V_{ch}-\frac{Q_{sc}}{2C_{ox}}+U_{T}\ln\left(\frac{N_D}{ni}\right)\\
	&+U_{T}\ln\left(1+\frac{Q_{sc}^2}{\theta}\right),
	\end{split}
	\end{equation}
	where $ \theta=8\epsilon_{si}qN_DU_T$.
	\subsubsection{Drain Current}
	The relationships derived previously (\ref{10}) and (\ref{12}) for depletion and accumulation modes give rise to explicit relationships for the channel current \cite{sallese2011charge}. The drain current in depletion is given by
	\begin{equation} \label {14}
	\begin{split}
	&I_{Dep}=\!\mu\frac{W}{L_g}\Biggl[\!\left(\frac{1}{8C_{sc}}-\frac{1}{4C_{ox}}\right)\!Q_{sc}^2\!-\!\frac{Q_{sc}^3}{12Q_fC_{sc}}\\
	&+\left(\frac{Q_f}{2C_{ox}}+2U_T\right)Q_{sc}-U_TQ_f\ln\left(1+\frac{Q_{sc}}{Q_f}\right)^2\Biggr]^D_S,
	\end{split}
	\end{equation}
	and in accumulation we have
	\begin{equation} \label {15}
	\begin{split}
	&I_{Acc}=\!\mu\frac{W}{L_g}\Biggl[\left(\frac{Q_f}{2C_{ox}}+2U_T\right)Q_{sc}-\frac{1}{4C_{ox}}Q_{sc}^2\\
	&-U_TQ_f\ln\left(1+\frac{Q_{sc}^2}{8Q_fC_{sc}U_T}\right)\\
	&-2U_T\sqrt{8Q_fC_{sc}U_T}\arctan\left(\frac{Q_{sc}}{\sqrt{8Q_fC_{sc}U_T}}\right)\Biggr]^D_S,
	\end{split}
	\end{equation}
	where $ \mu$ is the free carrier mobility assumed constant along the channel in this work. Also, for gate voltages where a hybrid channel takes place \cite{sallese2011charge}, i.e. part of the channel (near the source) in accumulation and the rest in depletion, the drain current becomes
	\begin{equation} \label {155}
	I_{hyb}=I_{Acc}\Big|_S^{FB}+I_{Dep}\Big|_{FB}^D
	\end{equation}
	
	\begin{figure}[t]
		\centering
		\includegraphics[width=1\columnwidth]{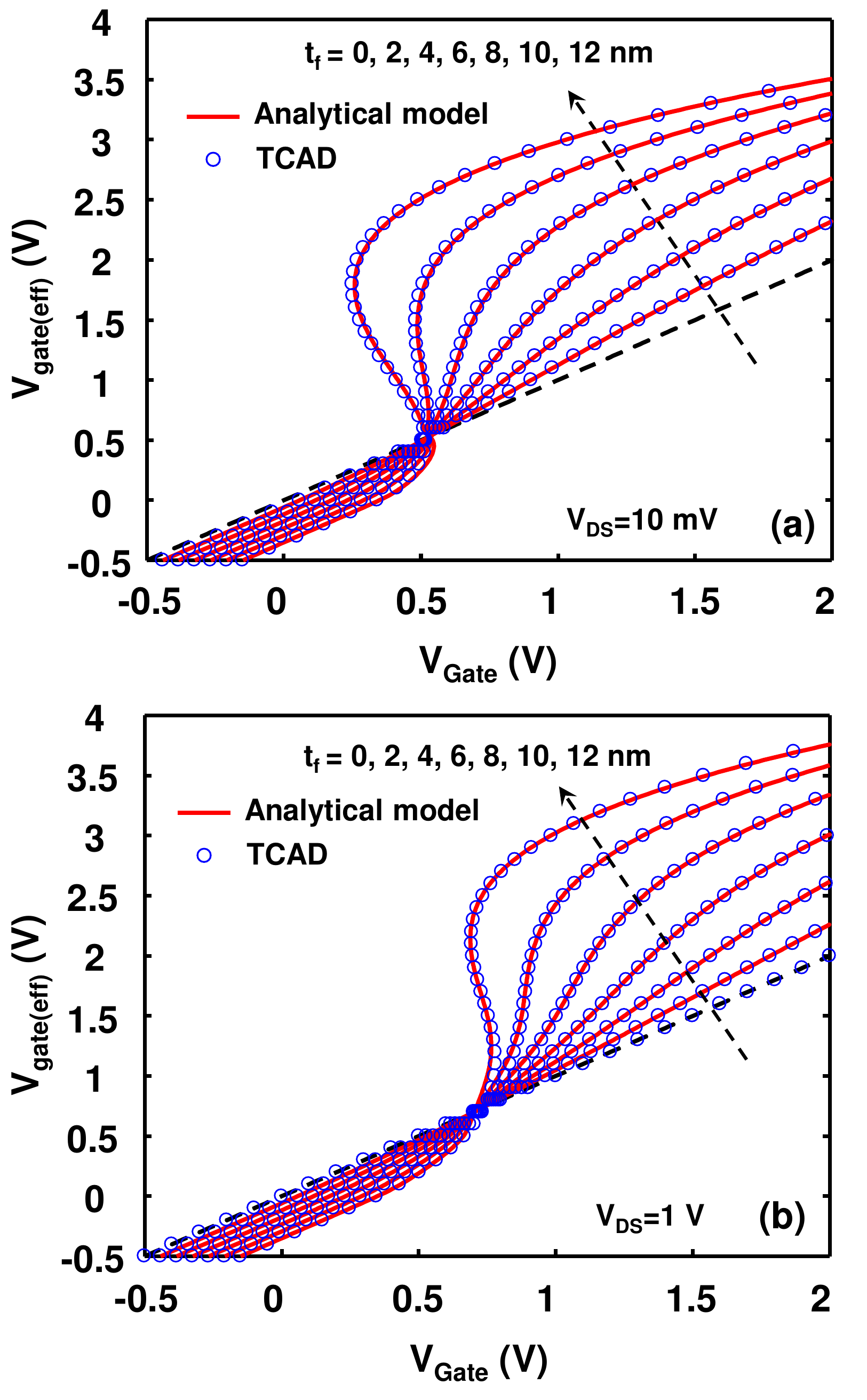}
		\caption{The potential of middle metal versus the applied gate voltage from the analytical model (lines) and TCAD simulations (circles) for the various thickness of ferroelectric from \SI{2}{nm} to \SI{12}{nm} (a) at $ V_{DS}$ = \SI{10}{mV} and (b) $ V_{DS}$ = \SI{1}{V}. By increasing the thickness of ferroelectric the voltage amplification increases.}
		\label{Fig2}\vspace{-0.5cm}
	\end{figure}
	\begin{figure}[t]
		\centering
		\includegraphics[width=1\columnwidth]{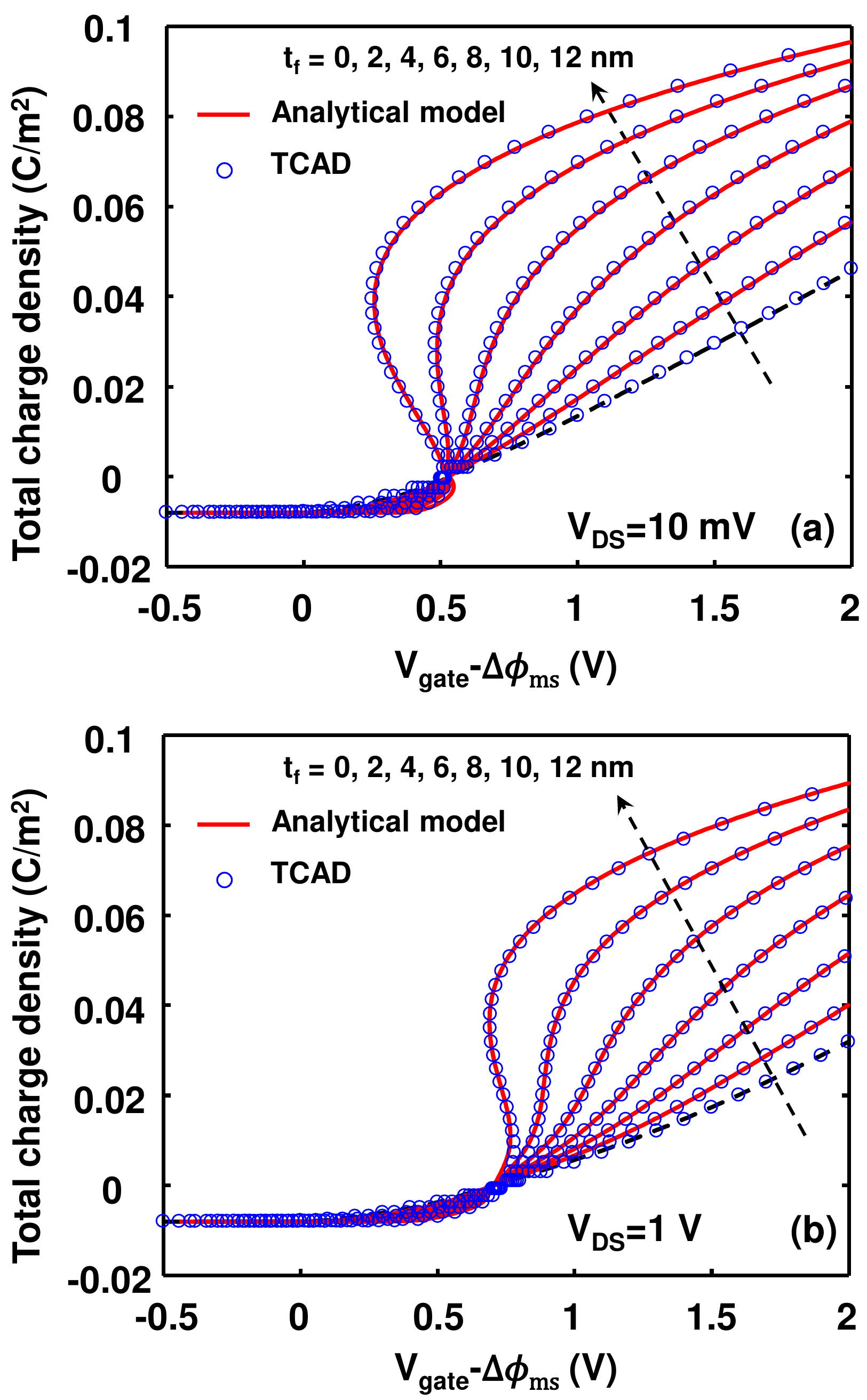}
		\caption{The total charge density in the ferroelectric versus the applied gate voltage from the analytical model (lines) and TCAD simulations (circles) for the various thickness of ferroelectric from \SI{2}{nm} to \SI{12}{nm} (a) at $ V_{DS}$ = \SI{10}{mV} and (b) $ V_{DS}$ = \SI{1}{V}.}
		\label{Fig3}\vspace{-0.5cm}
	\end{figure}
\vspace{-0.2cm}
	\subsection{Merging JLFET with Ferroelectric}
	\subsubsection{Landau Equation}
	In this section, we will merge the model of the JLFET described above with the core relation governing the ferroelectric layer. 
	\subsubsection{Total Charge Density}
	To obtain $V_f$ from (\ref{4}) the total charge density in the ferroelectric material $Q$ must be known. This is obtained by calculating the integral of the semiconductor charge density over the channel length
	\begin{equation} \label {17}
	WL_g\times 2Q=-\int_{0}^{L_g}WQ_{sc}dx.
	\end{equation}
	The total charge density is the sum of fixed and mobile charges $Q_{sc}=Q_f+Q_m$. Hence, we can write
	\begin{equation} \label {18}
	L_g\times 2Q=-Q_fL_g-\int_{0}^{L_g}Q_{m}dx.
	\end{equation}
	Although we do not know how $Q_m$ is related to $x$, we know the relation between $Q_m$ and $V_{ch}$ from (\ref{10}) and (\ref{12}) in depletion and accumulation modes respectively. In addition, from the drain current $I_{ds}$ relationship, $dx$ and $V_{ch}$ are linked as follows
	\begin{equation} \label {19}
	dx=-\frac{\mu Q_m}{I_{ds}}dV_{ch}.
	\end{equation}
	By substituting (\ref{19}) in (\ref{18}) the integral in space turns into an integral over the potential of the channel from the source (S) to the drain (D)
	\begin{equation} \label {20}
	Q=-\frac{Q_f}{2}+\frac{\mu}{2L_gI_{ds}}\int_{S}^{D}Q_{m}^2dV_{ch}.
	\end{equation}
	In depletion mode, $dV_{ch}$ is obtained from (\ref{10}) as follows
	\begin{equation} \label {21}
	dV_{ch}=\left(\frac{1}{2C_{ox}}+\frac{2U_TQ_{sc}}{Q_f^2-Q_{sc}^2}+\frac{Q_{sc}}{4C_{sc}Q_f}\right)dQ_{sc}.
	\end{equation}
	Since $dQ_{sc}=dQ_m$, we introduce (\ref{21}) in (\ref{20}). After solving the integral, an analytical and explicit relation for the total charge in the ferroelectric is obtained (in depletion mode):
	\begin{equation} \label {23}
	\begin{split}
	&Q=-\frac{Q_f}{2}+\frac{\mu}{2L_gI_{ds}}\biggl\{\frac{Q_m^3}{6C_{ox}}\\
	&+2U_T\biggl[Q_mQ_f-\frac{Q_m^2}{2}-2Q_f^2\ln(2Q_f+Q_m)\biggr]\\
	&+\frac{Q_m^4}{16C_{sc}Q_F}+\frac{Q_m^3}{12C_{sc}}\biggr\}^D_S.
	\end{split}
	\end{equation}
	
	In accumulation, we have
	\begin{equation} \label {24}
	dV_{ch}=\left(\frac{1}{2C_{ox}}+\frac{2U_TQ_{sc}}{Q_{sc}^2+\theta}\right)dQ_{sc}.
	\end{equation}
	Substituting (\ref{24}) in (\ref{20}) and then solving the integral gives an analytical and explicit relationship for the total charge density of the ferroelectric when the whole device is biased in accumulation:
	\begin{equation} \label {26}
	\begin{split}
	&Q=-\frac{Q_f}{2}+\frac{\mu}{2L_gI_{ds}}\biggl\{\frac{Q_m^3}{6C_{ox}}\\
	&-2U_T\biggl[\frac{Q_m^2}{2}+\frac{Q_f^2-\theta}{2}\ln(Q_{sc}^2+\theta)-Q_mQ_f\\
	&+2Q_f\sqrt{\theta}\arctan\big(\frac{Q_{sc}}{\sqrt{\theta}}\big)\biggr]\biggr\}^D_S.
	\end{split}
	\end{equation}
	\begin{figure}[t]
		\centering
		\includegraphics[width=1\columnwidth]{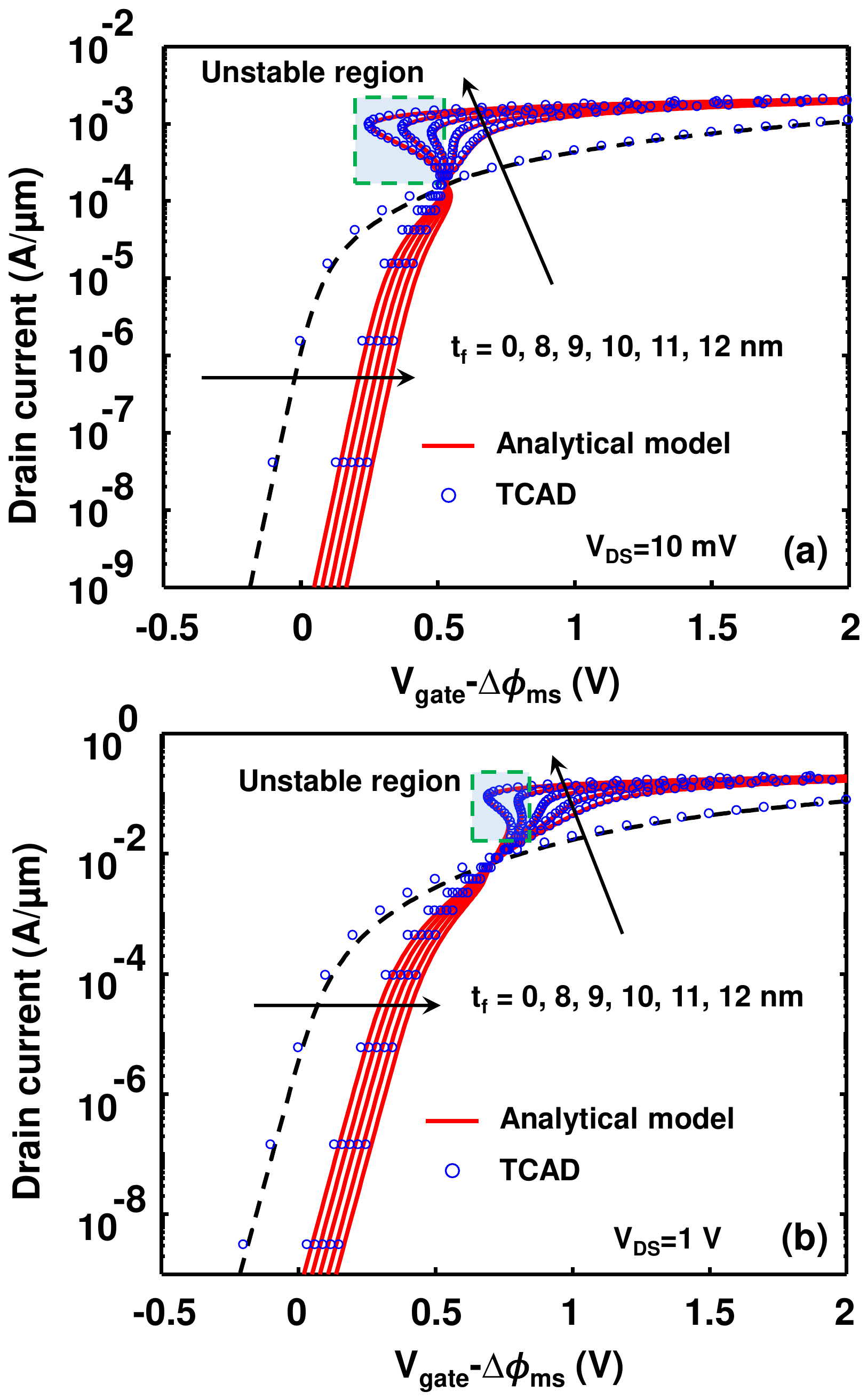}
		\caption{Drain current versus the applied gate voltage from the analytical model (lines) and TCAD simulations (circles) for the various thickness of ferroelectric from \SI{8}{nm} to \SI{12}{nm} (a) at $ V_{DS}$ = \SI{10}{mV} and (b) $ V_{DS}$ = \SI{1}{V}.}
		\label{Fig4}\vspace{-0.5cm}
	\end{figure}

	The way charges and voltages have been calculated is now explained: for a given effective gate voltage $V_{gate(eff)}$, the total charge density of ferroelectric $Q$ is known either from (\ref{23}) or (\ref{26}). Next, introducing $Q$ in (\ref{4}) and using equation (\ref{1}), the applied gate voltage $V_{gate}$ is finally obtained.
		
	Proceeding through this sequence greatly simplifies the way plots are obtained, avoiding cumbersome self-consistent calculations, and providing a clear and simple understanding of the imbrication of the different parts of the model.
		
	To confirm the validity of the model, we performed simulations with TCAD software. The simulation result of NCDG JLFET is numerically calculated by combining SILVACO TCAD software with a MATLAB script to include the Landau equation.
		
	For extracting the coefficients in relation (\ref{4}), we rely on the data published in literature \cite{hoffmann2019unveiling} where authors fabricated and characterized metal–ferroelectric–metal (MFM) capacitors by using an HZO film as a ferroelectric material in  \SI{11.6}{nm} thickness. They extracted a $P$-$E$ hysteresis loop with a remanent polarization $P_r$ of about \SI{17}{\micro C/cm^{2}} and a coercive field $E_c$ of about \SI{1.2}{MV/cm}. By using the experimental data of remanent polarization and coercive field and based on the approach presented in \cite{ricinschi1998analysis}, we extracted the Landau coefficients. The effect of the film thickness on the ferroelectricity of an HZO film has been studied in \cite{hyuk2014effects}. The coercive electric field of the HZO film is almost constant in various thicknesses. However, its remnant polarization decreases when the film thickness increases. This effect has been attributed to the change in the grain size of the HZO film during the atomic layer deposition (ALD) process \cite{hyuk2014effects}. Unlike HZO and hafnia films with other dopants, Gd:HfO\textsubscript{2} shows no reduction of the polarization in a different range of ferroelectric material \cite{hoffmann2015stabilizing}. Nevertheless, this effect merely changes the Landau coefficients for the different thicknesses of the ferroelectric film but the model can still predict the NC effect properly with the new coefficients. Also, for such a thin HZO film and high applied gate voltage, charge trapping effects might observe \cite{hoffmann2019unveiling}. However, we have proposed a core model and one can include this effect by employing the approach which is developed in our previous work \cite{rassekh2019modeling}.
		
	Next , we consider a double gate JLFET with \SI{100}{nm} channel length, \SI{1}{\micro m} channel width, and with \SI{10}{nm} of silicon thickness. The doping density and oxide thickness were set respectively to $ N_D$ = \SI{E19}{cm^{-3}} and \SI{1}{nm} (see Table \ref{tbl1}).

	Fig. \ref{Fig2}.a and b show the applied gate voltage versus the effective gate voltage for various thicknesses of the ferroelectric ranging from \SI{2}{nm} to \SI{12}{nm}, both at low and high $V_{DS}$ respectively. Lines and blue circles have been used for the analytical model and TCAD simulations, respectively. The analytical model at both low and high $V_{DS}$ demonstrates a full agreement with TCAD simulations. It can be seen that increasing the thickness of the ferroelectric layer increases the voltage amplification, which is the signature of the negative capacitance effect.
		\begin{figure*}[t]
		\centering
		\includegraphics[width=2\columnwidth]{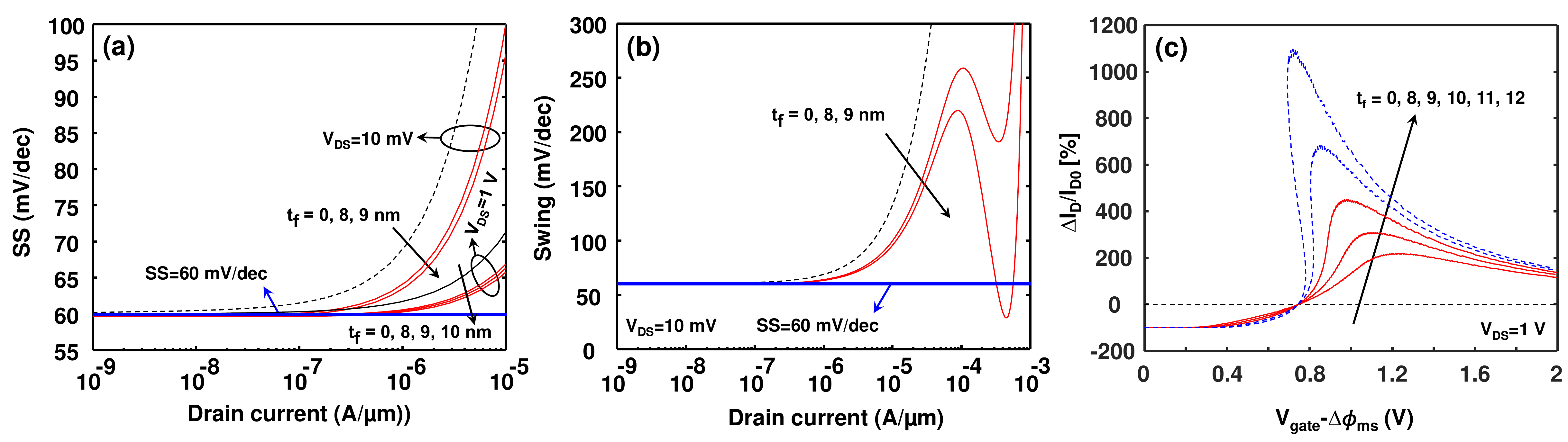}
		\caption{(a) SS versus the drain current in different thicknesses of ferroelectric for the various thickness of ferroelectric. (b) Swing versus the drain current for the various thickness of ferroelectric at $ V_{DS}$ = \SI{10}{mV}. (c) The percentage of increment of drain current ($\Delta I_D/I_{D0}$) versus applied gate voltage for the various thickness of ferroelectric from \SI{8}{nm} to \SI{12}{nm}. Thicknesses of ferroelectric in which NCDG JLFET operates in the hysteretic regime have been shown with dash lines.}
		\label{Fig6}\vspace{-0.6cm}
	\end{figure*}
	The total charge density of the ferroelectric versus the applied gate voltage at $V_{DS}$ = \SI{10}{mV} and $ V_{DS}$ = \SI{1}{V} obtained from TCAD simulations and from the model is plotted in Fig. \ref{Fig3}.a and Fig. \ref{Fig3}.b. To summarize, the total charge density of ferroelectric is now known in all regions of operation from depletion (\ref{23}) to accumulation (\ref{26}), resulting in a relationship between $V_{gate(eff)}$ and $V_{gate}$. Furthermore, we know the link between $V_{gate(eff)}$ and the mobile charge density in the channel, and so the relationship between $V_{gate}$ and $Q_{m}$, giving finally the expected $I_D-V_{gate}$ dependence.
	
	Fig. \ref{Fig4}.a and b illustrate the drain current versus the applied gate voltage at $V_{DS}$ = \SI{10}{mV} and $ V_{DS}$ = \SI{1}{V} for ferroelectric thicknesses from \SI{8}{nm} to \SI{12}{nm}. Increasing the thickness of ferroelectric causes a shift to the higher voltage in the subthreshold region, because in subthreshold the amount of fixed charge in the channel dominates over the mobile charge ($-Q_f/2$ in equation (\ref{23})). This makes a constant background of charge density regarding to the $t_f$ in the Landau equation. Beyond a threshold voltage, the slope increases and above a critical thickness $t_{cr}$ a snaps back due to the hysteresis effect of a ferroelectric layer is observed. These regions are unstable which are depicted with a rectangular in Fig. \ref{Fig4}. In order to have a non-hysteretic operation, the total gate capacitance should be positive in the whole range of the gate voltage \cite{saeidi2016double}. Fig. \ref{Fig6}.a shows SS versus the drain current in different thicknesses of ferroelectric at low and high $V_{DS}$. It illustrates that NC systematically reduces SS, even if not sub-\SI{60}{mV/dec} in all regimes. Subthermionic values down to sub \SI{30}{mV/dec} are achieved at moderate drain currents and low $V_{DS}$ (see Fig. \ref{Fig6}.b). It means that NC still causes a significant improvement in the overdrive voltage. On the other hand, by using a negative capacitance, we gain a remarkable increase in ON current as illustrated in Fig. \ref{Fig6}.c.  Fig. \ref{Fig6}.c shows the percentage of increment of drain current versus applied gate voltage in different thicknesses of the ferroelectric layer.
	
	Our results are conceptually in agreement with the experimental and simulation works \cite{saeidi2016double,saeidi2017negative}. But it is worth noting that these are different from previous work on modeling and simulation of a junctionless transistor with ferroelectric \cite{mehta2016modeling} where no such behavior was evidenced.
	\vspace{-0.2cm}
	\section{Short Channel Effect}
	In this section, we investigate how negative capacitance can affect the DIBL short channel effect. To this purpose, we need to solve the two-dimensional Poisson-Boltzmann relationship in the channel in the subthreshold. This was done for a regular JLFET in \cite{jazaeri2013analytical}.

	The potential distribution obtained from the 2D Poisson-Boltzmann relation is expressed as follows \cite{jazaeri2013analytical}:
	\begin{equation} \label {27}
	\begin{split}
	&\psi(x,y)=\psi_s(x)\left[1+y(1-\frac{y}{T_{sc}})\frac{C_{ox}}{\epsilon_{si}}\right]\\
	&+(\Delta \phi_{ms}-V_{gate(eff)})y(1-\frac{y}{T_{sc}})\frac{C_{ox}}{\epsilon_{si}},
	\end{split}
	\end{equation}
	where $\psi_{s}(x)$ is the surface potential at $y=0$ and $y=T_{sc}$ which is defined as follows
	\begin{equation} \label {28}
	\psi_s(x)=\eta \exp(\delta x)+\zeta \exp(-\delta x)+\lambda,
	\end{equation}
	\begin{equation} \label {281}
	\delta=\sqrt{\frac{2C_{ox}}{\epsilon_{si}T_{sc}}},
	\end{equation}
	\begin{equation} \label {282}
	\lambda=V_{gate(eff)}-\Delta \phi_{ms}+\frac{qN_D}{\delta^2\epsilon_{si}},
	\end{equation}
	\begin{equation} \label {283}
	\zeta=\frac{-\lambda\left[\exp(\delta L_g)-1\right]-V_{DS}}{2\sinh(\delta L_g)},
	\end{equation}
	\begin{equation} \label {284}
	\eta=-\zeta-\lambda,
	\end{equation}
	\begin{figure*}[t]
		\centering
		\includegraphics[width=2\columnwidth]{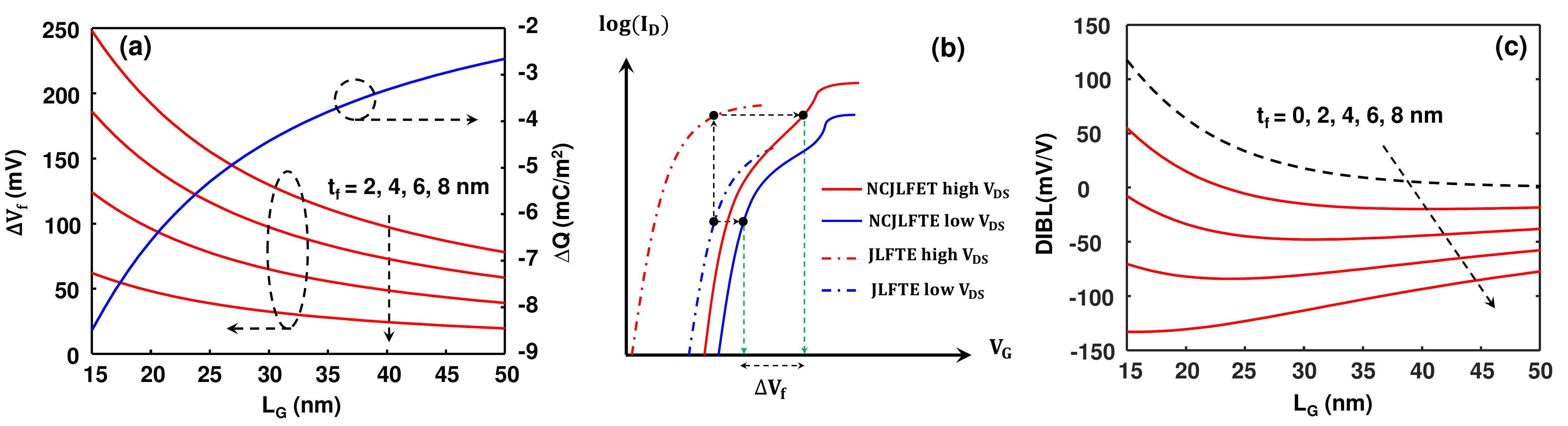}
		\caption{(a) The difference of the potential across the ferroelectric (left axis) and the difference of total charge density of ferroelectric (right axis) in high $ V_{DS}$ and low $ V_{DS}$ versus the channel length. $ \Delta V_{f}$ somehow represents $ \Delta V_{G}$. The model confirms that $ \Delta V_{f}$ and the absolute value of $ \Delta Q$ increase by going to the shorter channel length.(b) The schematic of the $I$-$V$ characteristic of a regular double gate JLFET and a double gate JLFET with negative capacitance at low and high $V_{DS}$. (c) DIBL of an NCDG JLFET in different thicknesses of ferroelectric versus the channel length.}
		\label{Fig5}\vspace{-0.5cm}
	\end{figure*}

	 By introducing the surface potential from equation (\ref{28}) in (\ref{8}), the total charge density in the channel (which is a function of $V_{DS}$) becomes
	\begin{equation} \label {29}
	\begin{split}
	Q_{sc}=&-2C_{ox}\biggl\{V_{gate(eff)}-\Delta\phi_{ms}\\
	&-\left[\eta \exp(\delta x)+\zeta \exp(-\delta x)+\lambda\right]\biggr\},
	\end{split}
	\end{equation}
	Finally, the total charge density in the ferroelectric can be obtained by substituting (\ref{29}) in (\ref{17}) and calculating the integral along the lateral direction of the channel from $0$ to $L_g$: 
	\begin{equation} \label {30}
	\begin{split}
	&Q=\frac{C_{ox}}{L_g}\biggl\{(V_{gate(eff)}-\Delta\phi_{ms}-\lambda)L_g\\
	&-\left[\frac{\eta}{\delta}\exp(\delta L_g)-\frac{\zeta}{\delta}\exp(-\delta L_g)+\frac{\zeta-\eta}{\delta}\right]\biggr\}.
	\end{split}
	\end{equation}
	Since $Q$ is a function of the drain voltage, we can estimate how much $V_{DS}$ affects the charge density of ferroelectric. Therefore, we define $\Delta Q=Q_h-Q_l$ as the difference in the charge density of ferroelectric between high and low $V_{DS}$:
	\begin{equation} \label {31}
	\Delta Q=\frac{-2C_{ox}\Delta V_{DS}}{2L_g\delta \sinh(\delta L_g)}\left[\cosh(\delta L_g)-1\right],
	\end{equation}
	where $\Delta V_{DS}=V_{DS(high)}-V_{DS(low)}$.
	
	To find $\Delta V_f$, the difference of the electrostatic potential across the ferroelectric at high and low $V_{DS}$, we neglect the coefficient $\gamma$. Hence, $\Delta V_f$ becomes
	\begin{equation} \label {32}
	\Delta V_f=(2\alpha t_fQ_h+4\beta t_fQ_h^3)-(2\alpha t_fQ_l+4\beta t_fQ_l^3),
	\end{equation}
	which can be further simplified
	\begin{equation} \label {33}
	\Delta V_f=2\alpha t_f\Delta Q+4\beta t_f\Delta Q^3+12\beta t_f\Delta Q^3Q_hQ_l.
	\end{equation}
	 It happens that $Q_h$ and $Q_l$ are the only contributions to $\Delta V_f$ that depend on the effective gate voltage. Still, it happens that the last term in (\ref{33}) containing these quantities is negligible, meaning that we can approximate $\Delta V_f$ as follows
	\begin{equation} \label {34}
	\Delta V_f\approx 2\alpha t_f\Delta Q+4\beta t_f\Delta Q^3.
	\end{equation}
	
	Fig. \ref{Fig5}.a depicts $\Delta V_f$ and $\Delta Q$ versus the channel length on the left and right axis respectively. We see that the absolute value of $\Delta Q$ increases by going to the shorter channel lengths, and as a result, $\Delta V_f$ increases as well. This is an advantage because it predicts that negative capacitance mitigates short channel effect at high $V_{DS}$. Actually $ \Delta V_{f}$ represents somehow $ \Delta V_{G}$, i.e. the difference of $ V_{G}$ between high and low $ V_{DS}$. 
	
	The schematic drawing in Fig. \ref{Fig5}.b illustrates this effect. It compares the $I$-$V$ characteristic of a regular double gate JLFET and a double gate JLFET in presence of negative capacitance at low and high $V_{DS}$. As we mentioned in the previous section, a negative capacitance causes a potential amplification as much as $V_f$. Although $V_f$ is almost the same for high and low $V_{DS}$ for the long channel device, it becomes less effective at low $V_{DS}$ in regard to high $V_{DS}$ for a short channel device. In fact, in presence of short channel effect, the $I$-$V$ characteristic that shifts towards high $V_{DS}$ is compensated by $\Delta V_f$ in NCDG JLFET, thus DIBL reduction in a specific thickness of the ferroelectric film would be feasible. 
	
	Unlike conventional MOSFETs where DIBL is related to the shift in the surface potential since most of the current flow from the Si-SiO{\textsubscript{2}} interface, in JLFETs when they operate below the threshold, the interface no longer represents the lowes energy across the channel and the DIBL must be calculated from the center potential shift upon the drain voltage as follows \cite{jazaeri_sallese_2018,jazaeri2013analytical}
	\begin{equation} \label {38}
	DIBL_{\Delta\psi}=\frac{\Delta\psi_{BCP,min}}{\Delta V_{DS}},
	\end{equation}
	where $\Delta\psi_{BCP,min}$ is the difference of minimum body center potential in low and high $V_{DS}$. We know from \cite{jazaeri_sallese_2018,jazaeri2013analytical} that the body center potential is linked to the surface potential:
	\begin{equation} \label {39}
	\psi_{BCP}=a\psi_{s}+b,
	\end{equation}
	where $a$ and $b$ coefficients are given by
	\begin{equation} \label {40}
	a=1+\frac{1}{8}\delta^2T_{sc}^2,
	\end{equation}
	\begin{equation} \label {41}
	b=(a-1)(\Delta\phi_{ms}-V_{gate}+V_f).
	\end{equation}
	Therefore, the difference of body center potential of an NCJLFET in low and high $V_{DS}$ will be
	\begin{equation} \label {42}
	\Delta \psi_{BCP}=a\Delta \psi_{s}+(a-1)\Delta V_f.
	\end{equation}
	Then $\Delta \psi_{s}$ obtains from (\ref{8}) as follows
	\begin{equation} \label {43}
	\Delta \psi_{s}=-\Delta V_f+\frac{\Delta Q_{sc}}{2C_{ox}}.
	\end{equation}
	We can replace $\Delta Q_{sc}$ with $\Delta Q_{sc_{JL}}+\Delta Q^{\prime}_{sc}$ where $\Delta Q_{sc_{JL}}$ is the difference of the total charge density in the semiconductor in a normal JLFET without NC. Hence, $\Delta \psi_{s}$ links to the difference of the surface potential of a normal JLFET in low and high $V_{DS}$ ($\Delta \psi_{s_{JL}}=\Delta Q_{sc_{JL}}/2C_{ox}$) as follows 
	\begin{equation} \label {44}
	\Delta \psi_{s}=\Delta \psi_{s_{JL}}-\Delta V_f+\frac{\Delta Q^{\prime}_{sc}}{2C_{ox}}.
	\end{equation}
	If we neglect $\Delta Q^{\prime}_{sc}/2C_{ox}$ and introduce (\ref{44}) in (\ref{42}), we will have (we know that $\Delta \psi_{BCP_{JL}}=a\Delta \psi_{s_{JL}}$)
	\begin{equation} \label {45}
	\Delta \psi_{BCP}\approx\Delta \psi_{BCP_{JL}}-\Delta V_f.
	\end{equation}
	Therefore, DIBL in NCJLFET shifts as much as $\Delta V_f/\Delta V_{DS}$ respect to the DIBL in a JLFET.
	\begin{equation} \label {46}
	DIBL_{\Delta\psi}\approx DIBL_{\Delta\psi_{JL}}-\frac{\Delta V_f}{\Delta V_{DS}},
	\end{equation}
	where $DIBL_{\Delta\psi_{JL}}$ defines as follows \cite{jazaeri_sallese_2018,jazaeri2013analytical}
	\begin{equation} \label {47}
	DIBL_{\Delta\psi_{JL}}=\frac{2a\sqrt{\eta^{\prime}\zeta^{^{\prime}}}}{\Delta V_{DS}}\bigg(\sqrt{1+\frac{\Delta}{\eta^{\prime}}-\frac{\Delta}{\zeta^{\prime}}-\frac{\Delta^2}{\eta^{\prime}\zeta^{\prime}}}-1\bigg),
	\end{equation}
	\begin{equation} \label {383}
	\zeta^{\prime}=\frac{(-b-a\lambda)\left[\exp(\delta L_g)-1\right]-V_{DS}}{2a\sinh(\delta L_g)},
	\end{equation}
	\begin{equation} \label {384}
	\eta=-\frac{b}{a}-\zeta^{\prime}-\lambda,
	\end{equation}
	\begin{equation} \label {50}
	\Delta=\frac{\Delta V_{DS}}{2a\sinh(\delta L_g)}.
	\end{equation}
	
	Fig. \ref{Fig5}.c shows DIBL of an NCDG JLFET in different thicknesses of ferroelectric versus the channel length. We see that the value of DIBL decreases by increasing the thickness of the ferroelectric. As expected, in the short channel lengths, the different amount of potential drops across the ferroelectric film at low and high $V_{DS}$ which causes a reduction in the DIBL even to the negative values. Hence, it appears that by choosing a proper thickness for the ferroelectric film, an improvement in the DIBL would be achievable e.g. in our case $t_f$$\approx$\SI{3}{nm} gives the lowest DIBL. Further studies for an optimization DIBL strategy and trade-offs with other performance figures of merit are needed, which was beyond the scope of this paper.
	\vspace{-0.2cm}
	\section{Conclusion}
	An analytical charge-based model for symmetric double-gate junctionless FETs with negative capacitance was developed. The model incorporates the impact of the negative capacitance of ferroelectric on DC electrical characteristics of double gate JLFETs by proposing an analytical and explicit equation for the total charge density of ferroelectric in depletion and accumulation modes. The model confirms that using the negative capacitance in junctionless transistors means that the gate overdrive voltage decreases, which can be interpreted as lower energy consumption. The model also shows that the subthreshold slope almost remains constant in NCDG JLFET, but an improvement of swing for above the threshold causes a significant enhancement in ON current. In addition, an analysis of the short channel effect predicts an improvement when negative capacitance is observed. The model has been compared to TCAD simulations with an excellent agreement in all regions of operation from deep depletion to accumulation and linear to saturation.
\vspace{-0.2cm}
	\section{APPENDIX}
	The total polarization $P$ can be expressed as the sum of a linear and switching dipole $P_D$ contributions as follows \cite{sallese2004switch}
	\begin{equation} \label {35}
	P=\epsilon_{0} \chi E+P_D.
	\end{equation}
	Hence, the surface charge density $\sigma$ of ferroelectric becomes
	\begin{equation} \label {36}
	\sigma=V\frac{\epsilon_{f}}{t_f}+P_D,
	\end{equation}
	where $\epsilon_{f}=\epsilon_{0}\left(1+\chi\right)$. Thus the capacitance becomes
	\begin{equation} \label {37}
	C=\frac{d\sigma}{dV}=\frac{1}{t_f}\left(\epsilon_{f}+t_f\frac{dP_D}{dV}\right).
	\end{equation}
	We can replace $dV$ with $t_fdE$ and then $dP_D/dE=dP/dE-\epsilon_{0} \chi$. Therefore, the capacitance is related to the slope of the $P$-$E$ curve as expressed in (\ref{222}).
	\vspace{-0.2cm}
	\section*{Acknowledgement}
	This work was financially supported by the European Research Council (ERC) under the ERC Advanced Grants Milli-Tech (695459 ERC Millitech).
	
	\bibliographystyle{IEEEtran}

\end{document}